# Quantum Internet, Governance, Trust, and the Promise of Secure Communication:

On building a Quantum Internet that will be used

Pieter E. Vermaas,
    Philosophy Section, TU Delft,
    Jaffalaan 5, 2628 BX Delft, the Netherlands
    p.e.vermaas@tudelft.nl
    *Corresponding Author*

Luca Possati
    Philosophy Section, University of Twente,
    Drienerlolaan 5, 7522 NB Enschede, the Netherlands

Zeki C. Seskir
    Institute for Technology Assessment and Systems Analysis, Karlsruhe Institute of Technology,
    Karlstraße 11, 76133 Karlsruhe, Germany

Abstract

The development of quantum technologies has been accelerating in the last decade, turning them into emerging technologies that need explicit attention by decision-makers at national funding agencies, companies and governments. In this paper we consider the governance of quantum internet, a new type of communication network developed for the promise that it can deliver inherently secure communication solely through its technical design. This paper gives a general analysis of the functions quantum internet offer, and then challenges this proposition by arguing that trust – an essential precondition for users to adopt this technology – cannot be guaranteed by technical features alone. Instead, trust is fundamentally a social phenomenon, shaped by how quantum internet is governed, operated, and regulated. Therefore, the ultimate success of quantum internet in fulfilling its promise of secure communication will depend not just on its technical hardware and software but also on the policies and frameworks for running these capacities, and on the public trust in those policies and frameworks. With this argument we arrive at recommendations to decision-makers for developing quantum internet and its governance in a way that quantum internet that can be trusted for its promised capacities.

Keywords:
Quantum internet, Communication security, Technology governance, Trust, Operators, Regulators, Quantum Key Distribution (QKD)

# 1. Introduction

The development of quantum technologies has been accelerating in the last decade, turning them from specialist topics in scientific research to emerging technologies that needs explicit attention by decision-makers at national funding agencies, companies, and governments. This attention is for a larger part focused on the guidance of research, innovation, and commercialization of quantum technologies, yet is increasingly widening to the responsible governance of the technologies and their applications. This widening is in part driven by multilateral political organizations (e.g., [1,2]), anticipating that quantum technologies were to be recognized by many states as technologies strategic to their security and autonomy. The call for responsible governance of quantum technologies has also reached academic research (e.g., [3,4]), and in this paper we contribute to this call by considering the governance of quantum internet.

Quantum internet is a communication network built with quantum technologies, and developed for the promise that it can deliver inherently secure communication solely through its technical design. We give a general analysis of the functions quantum internet offer, and then challenge this proposition by arguing that trust – an essential precondition for users to adopt this technology – cannot be guaranteed by technical features alone. Instead, trust is fundamentally a social phenomenon, shaped by how quantum internet is governed, operated, and regulated. Therefore, the ultimate success of quantum internet in fulfilling its promise of secure communication will depend not just on its technical hardware and software but also on the policies and frameworks for running these capacities, and on the public trust in those policies and frameworks. The insight that people need to trust new technologies and that the governance of the technologies is central for arriving at that trust, has reached academic results as well, as can be witnessed in this journal [5,6]. In this paper we build on this insight for arriving at recommendations to decision-makers for developing quantum internet and its governance in a way that quantum internet can be trusted for its promised capacities.

Quantum internet is a new type of communication network that relays information encoded in quantum-mechanical systems, called qubits, rather than information encoded digitally as is sent through the digital internet. Quantum internet consists technically of links, nodes, repeaters, and endpoints, and operates by layers of software [7-10]. Moreover, as with any technological infrastructure, it is a sociotechnical system constituted also by its users as well as its operators, regulators, their rules, their governance, and their institutions.

Since the beginning of the century the quantum internet was envisaged [11] and in the last decade its development accelerated [7], leading to the creation of the first quantum communication links with increasing sophistication. Impressive initial achievements such as the satellite quantum links between China and Austria and China and South Africa [12,13], the 2000-kilometer long quantum communication network in China [14], and the concerted buildup of quantum internet in Europe [9] have been achieved and are instilling confidence into the technical progress made in the field.

When compared to digital internet, quantum internet is estimated to bring a series of new functionalities and new applications such as providing intrinsically secure communication, linking quantum computers, and disseminating non-classical correlations over long distances. Specifically, the first application makes quantum internet an attractive infrastructure to have. Cybersecurity is a value that has gained prominence by the increased dependency of states, companies, and citizens on digital communication, the soaring rates of digital hacks and cyber attacks, and the threat of quantum computers [15,16]. However, in spite of this attractiveness, doubts have also been expressed about the security quantum internet can deliver, specifically from state actors (in the USA by [17]; in Germany by [18]). Building on these doubts, we argue in this paper that the possibility of using quantum internet for facilitating secure communication depends not only on availability but also on the willingness of users to actually adopt and use quantum internet. We take trust in quantum internet by users as a precondition to this willingness.

The idea that a certain level of trust is a precondition for utilizing a technology, and therefore also for the realization of its potential benefits, is of course not reserved for quantum technologies. For successful existing technologies that trust has been realized. Flying in a plane presupposes a sequence of levels of trust in the aircraft, in its crew, and in the system of traffic rules and safety regulations that apply to aviation. For new technologies that trust has to be acquired and may fail. Resistance to GMOs and reluctance towards cryptocurrencies can be explained in terms of distrusting the technology in question. With every novel technology we must build up trust in it, and it is not yet clear how that looks like for quantum internet. What are the elements that require trust, and what does it take for users to find that trust?

We take up these questions by looking at the promise that quantum internet can deliver intrinsically secure communication by its technical structure. In the next section we characterize quantum internet in general terms and introduce its main discussed functionalities. For developing our argument, we focus on one component of quantum internet in particular: the repeater. Repeaters boost the quantum-mechanical systems that carry the qubits through quantum internet and should do so without jeopardizing the functionality of quantum internet. The types of repeaters that currently exist lack that property by boosting qubits in a way that allows that the qubits are intercepted, copied, or manipulated. These types of repeaters are paradoxically called "trusted repeaters," which actually means that users should have to trust that during the boosting in the repeaters qubits are not actually intercepted, copied, or manipulated. New types of repeaters, called "quantum repeaters," that can boost qubits without the risk of interception, are imagined but are still in development. We argue that also with these quantum repeaters, users need to build up trust in the governance of quantum internet, in this case for believing that there are no attacks taking place on the hardware.

In Section 3 we spell out what trust may mean in the context of quantum internet. And in Section 4 we argue against the promise that quantum internet can deliver intrinsically secure communication by showing that trust in quantum internet, even with quantum repeaters, will depend on the way in which quantum internet is operated and regulated. Section 5 summarizes and gives a number of recommendations to decision-makers involved in the development of quantum internet and its governance on how users can find trust in quantum internet.

## 2. Quantum internet and its applications

### 2.1. Structure

When focusing on hardware, quantum internet consists of links, nodes, repeaters, and endpoints, that enable the sending of information in the form of qubits from one point to another. The quantum-mechanical systems that are sent are typically photons, or light particles, and usually the polarization of the photons represents the qubits. Links are channels through which the photons are sent and are typically glass fibers or corridors through open air or space. Nodes are connections between links and can be smaller quantum technological circuits or whole satellites. Repeaters are nodes that boost qubits, which is needed to move beyond the limited metropolitan level networks with diameters less than 50 km ([19], p. 84). Endpoints may be computers, sensors, or other machinery, and these endpoints may be digital or quantum technological devices.

In a quantum internet as envisaged, the core components are all quantum technological devices, connecting, say, quantum computers through quantum nodes, quantum links, and quantum repeaters (see Fig. 1). Devices count as quantum technological when their physical behavior is described by the laws of quantum mechanics,[1] making available typically quantum phenomena such as *superposition* of states of individual devices and *entanglement* between the states of different devices. Current quantum networks are however still a hybrid of quantum and digital technologies. For instance, endpoints can be digital computers and repeaters can be devices that just detect incoming qubits with digital measurement devices for creating new similar qubits that will be sent onwards (see Fig. 2). Devices count as digital when their physics is described by regular Newtonian physics or electromagnetics, which do not include the quantum phenomena of superposition and entanglement. The distinction is relevant because the functionality that is specific to quantum internet depends on superposition and entanglement. Hence, if for instance repeaters are digital devices that do not mediate superposition and entanglement, then part of the functionality of quantum internet is lost.

---

[1] Physical systems that are described by quantum mechanics are typically atomic systems, radiation and light, and systems that are isolated from interaction with their environment.

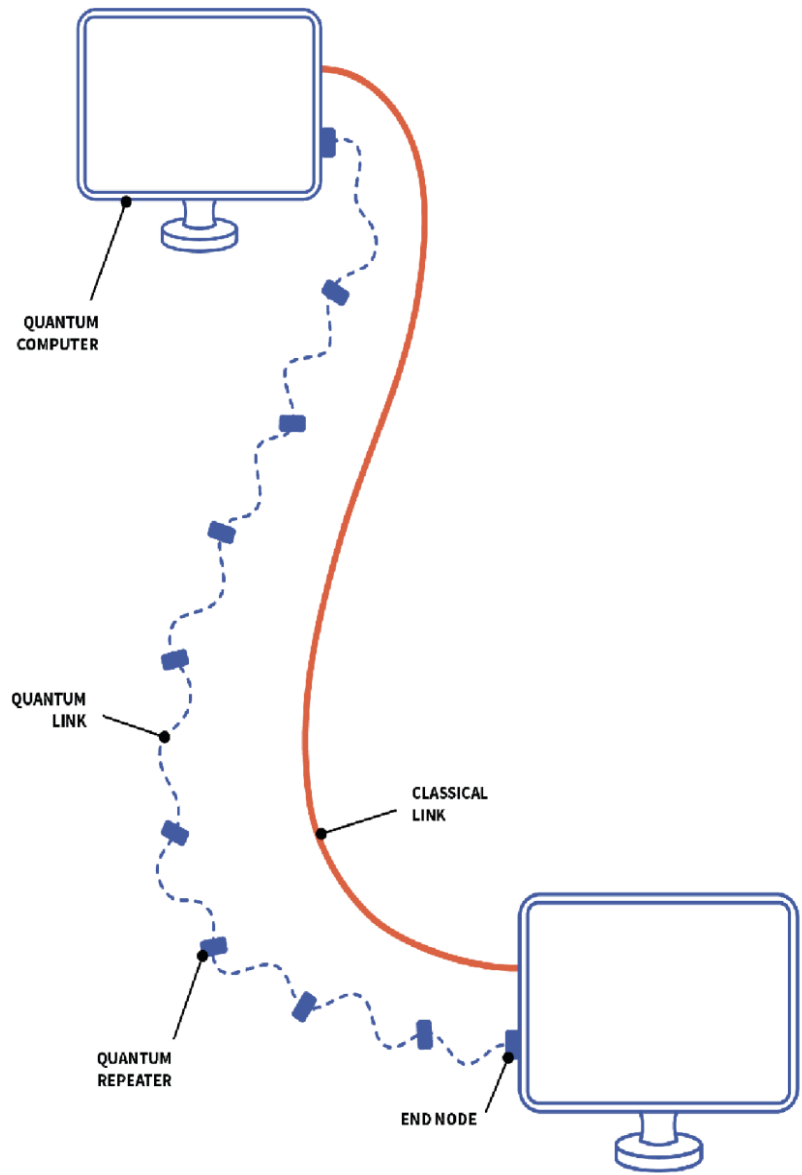

*Fig. 1 A fully quantum technological quantum internet; the blue components are quantum technological components; the red components are digital ones (source: [20])*

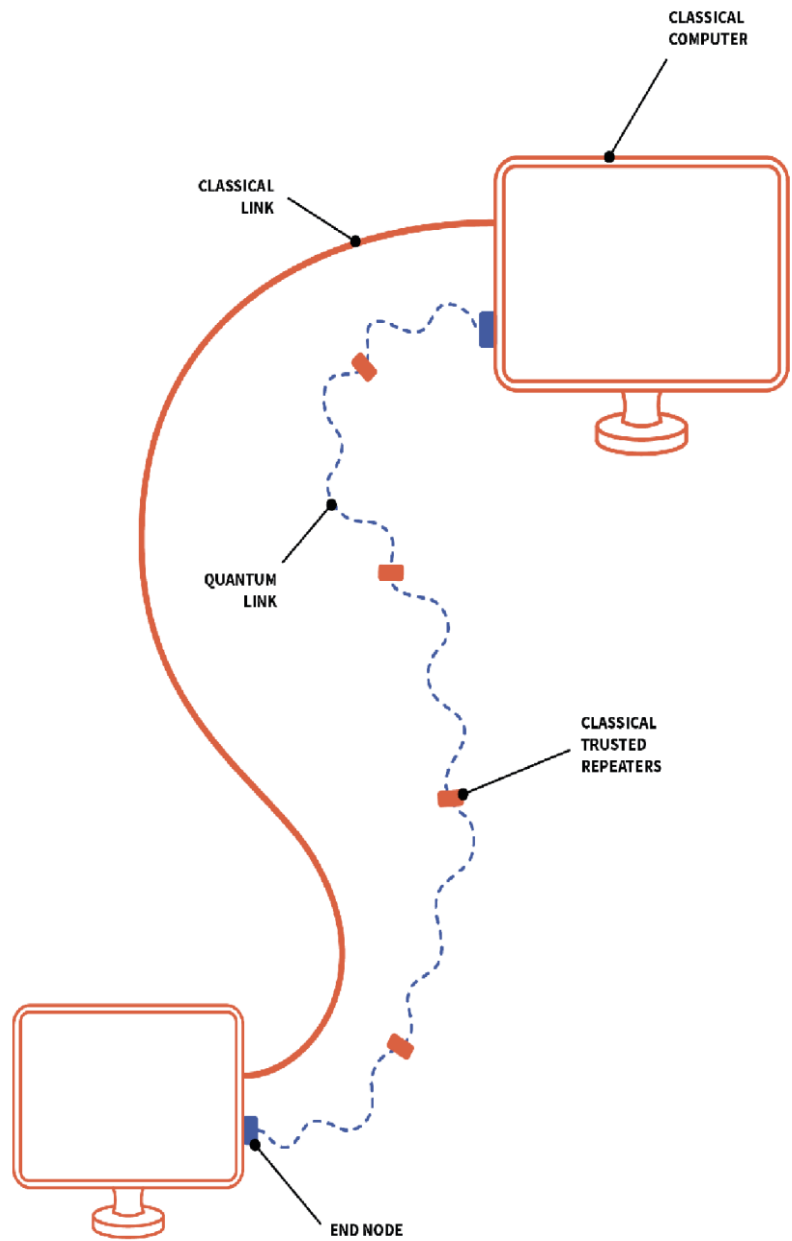

*Fig. 2 A quantum internet with digital components; the blue components are quantum technological components; the red components are digital ones (source: [20])*

The current inclusion of digital technology can be understood as a phase in the development of quantum internet ([7], which contains a more fine-grained description of the development phases of quantum internet). It can also be understood as a division of labor. Current quantum internet has a limited capacity of sending qubits, hence is typically not envisaged as a stand-alone communication network. Digital internet then remains the infrastructure for sending large volumes of data, and quantum internet can be used to exchange the cryptographic keys for encrypting that communication over the digital internet. Furthermore, for certain applications, such as quantum teleportation, existence of operational classical channels to send digital data are prerequisites.

## *2.2. Operation and regulation*

The operators of the classical internet include Internet Service Providers (ISPs), content providers, data center operators, and others who manage the physical infrastructure, including fiber-optic cables, servers, and routing systems. Major tech companies such as Google, Amazon, and Microsoft also play a key role by providing cloud services and content delivery networks. Regulators include international bodies like the Internet Corporation for Assigned Names and Numbers (ICANN), which oversees domain names and IP addresses, and the Internet Engineering Task Force (IETF), which develops technical protocols. Governments also regulate the internet through national policies on cybersecurity, data protection (e.g., GDPR in the EU), and censorship (e.g., FCC in the US and China's Great Firewall). Additionally, organizations like the International Telecommunication Union (ITU) set global communication standards, ensuring interoperability across different regions.

Quantum internet will similarly require coordination among network operators, service providers, policymakers, and researchers. Quantum internet relies on the classical internet to operate, and, on top of that, it also depends on entanglement distribution, quantum key exchange, repeater-based architectures, and similar quantum-mechanical applications which introduce new operational challenges [7]. Managing quantum networks will require protocols for entanglement routing, error correction, and interoperability between different quantum technologies [8]. These are complex and necessary protocols for quantum internet to run smoothly, hence regulating this new infrastructure – hardware, software, and operators – will be an essential issue to ensure security and standardization across the globe. Policymakers will need to establish clear rules for data protection, network governance, and international cooperation, as differences between national regulations may hinder the adoption of a global quantum internet significantly [9]. Similar to the existing internet, quantum internet will have its own ISPs, major industry players, national and international bodies, most of which are expected to be continuations of the existing regulatory infrastructure, such as ICANN and ITU. There is already a quantum internet research group (QIRG) under the IETF, an ICANN report titled Quantum Computing and the DNS [21], and the International Telecommunication Union [22] is working on standardization of quantum applications such as quantum key distribution (see below).

## 2.3. Functions

Quantum internet will offer some basic functionality that can be considered as novel in comparison to digital internet. These basic functions may seem less interesting themselves, yet when combined with other digital and quantum functions, novel applications can be created. Some of these applications are described in the next subsection; here we introduce two basic functions of quantum internet.

The first function is transmitting quantum-mechanical systems – typically photons – from one place A to another place B. The novelty of this transmission is that it may be intrinsically secure in the sense that the sender and receiver of the photons are able to determine if a third party is detecting the state of the sent photons. The mechanism that provides this security is quantum-mechanical, namely, detection of a sent photon by a third party will in general considerably change its state. The sender and receiver therefore can check if this change in state has taken place, and in this way discover if a third party has been detecting the photons.

The second function is distributing entangled quantum-mechanical systems – typically photons – to two (separate) places A and B. These entangled photons are in pairs created at a source S and then one photon of each pair is sent over quantum internet from S to place A and the second is set from S to B. The novelty of this function is that the entangled photons have a quantum-mechanical correlation that is independent of the distance between A and B, and that is stronger than correlations offered by digital means. Also this distribution of entanglement and correlations is intrinsically secure since detection of the state of one (or both) entangled photons changes the entanglement in a way that receivers at A and B can discover. Through quantum-mechanical operations, the number of entangled nodes can be increased beyond just two (A and B) to arbitrarily many, allowing applications that are not possible on digital internet.

## 2.4. Applications

The two functions that quantum internet offers have a number of interesting applications when combined with other functions of quantum-mechanical systems and digital apparatus, such as digital internet and quantum computers. We list some applications that are already identified in the literature, and it may be expected that more applications will be found in the future.

### Secure communication via qubit transmission

A first application is to use quantum internet for secure communication by sending qubits over quantum internet. When photons are the particles that are sent, then information can be coded into the spin property of these photons. The photons then become qubits, and can be sent from A to B. This application seems to make use of only the first basic function of quantum internet, yet more complex schemes are available too. For instance, the spin states of photons can also be sent from A to B through a "roundabout" scheme that does not require a direct link between A and B. By a scheme, which is called *quantum teleportation* [23], A and B share photons from entangled pairs created at a source S. The photon "*x*" whose state is to be transferred from A to B is in teleportation interacting at place A with one of the photons of the entangled pair. By this interaction, the other photon of the entangled pair, positioned at B, will inherit information about the initial state of the phone *x*, by which this state can be recovered at place B . Note that this transfer of the state of photon *x* from A to B takes place without there being a direct quantum internet link between A and B.

The communication by qubits is on both schemes intrinsically secure in the sense that sender and receiver can discover if an eavesdropper has interfered with either the sent qubit or, in case of teleportation, with the pair of entangled photons. However, neither of the schemes guarantee that secure communication can be established: the schemes allow sender and receiver to abort and delete communication in case eavesdropping occurs, yet do not prevent eavesdropping from occurring.

A second comment is that the volumes of information that can be exchanged over quantum internet by sending qubit over quantum internet will in the upcoming years not be comparable to the volumes that digital internet allows. A second application of quantum internet to enable secure communication over digital internet seems therefore more meaningful.

## Secure communication over digital internet: quantum key distribution

This second application of quantum internet is the exchange of secure encryption keys among senders and receivers who use digital internet for their communication. The advantage is one of division of labor: the security of communication over quantum internet is used for creating secure encryption keys, and the large capacity of the digital internet is used for sending the actual communication.

The exchange of encryption keys between A and B over quantum internet is called *quantum key distribution* (QKD), and can be done through sending qubits from A to B (called the BB84 protocol, referring to its publication by Bennett and Brassard in 1984 [24]) or through sharing pairs of entangled qubits between A and B (called the E91 protocol, after Ekert who published it in 1991 [25]). In both cases sender and receiver go first through a series of operations involving a batch of sent qubits, and a second exchange of a control sequence of bits over the digital internet. A and B can check by the control sequence whether a third party has been copying the qubits exchanged in the first batch. If that check reveals that no eavesdropping occurred, A and B can deduce a secure encryption key from the outcomes that the first set of operations on the qubits produced. They then can use that key for encrypting and decrypting information exchange over the digital internet.

### Remotely accessing quantum computers

A third application of quantum internet is to remotely access quantum computers. Quantum computers are another quantum technology developed at this moment for their ability to conduct computations practically not possible by digital (super) computers. Quantum computers do their computation with qubits, and quantum internet allows uploading the qubit input to calculations to a quantum computer from a place other than where the computer is located. Similarly, quantum internet allows downloading the output results. The exchange of input and output with the quantum computers is moreover again secure communication. It is also possible to carry out remote calculations on quantum computers via a scheme called *blind quantum computation* that hides the content of computation from third parties and even from the owners of the quantum computers [26].

### Distributed quantum computing

A fourth application of quantum internet consists of connecting different quantum computers to each other. Such a network of multiple connected quantum computers would enable distributed quantum computing, allowing the network to perform more complex calculations than would be possible with each single quantum computer [27]. Enhancing computational power of quantum computers is thus an important promise of quantum internet. Moreover, connecting quantum computers via a quantum internet is also a promising strategy to enhance the overall fault-tolerance, for a network of quantum computers may enable systematic detection and correction of errors, hence addressing error-sensitivity [7].

### Long-distance correlation

A fifth application of quantum internet is establishing correlations over long distances, with one key application being clock synchronization. Traditional synchronization methods rely on classical communication, which introduces time delays and inaccuracies over large distances. Quantum networks, however, can distribute entangled photon pairs to synchronize clocks with ultra-high precision, reducing errors caused by signal transmission delays [28]. This capability can play a crucial role in applications in fundamental physics, global positioning systems, and secure communication.

### Quantum sensing networks

A sixth application is using quantum internet as a sensing network by linking quantum sensors to it. Quantum sensors can detect minute changes in physical parameters like gravity, magnetic fields, or time with unprecedented accuracy [29]. By transmitting sensor data over a quantum internet, the information can be securely processed using quantum computers, enabling high-speed analysis for applications in fields such as geophysics, navigation, and fundamental physics research [7].

### Voting

A final application we will cover here is quantum voting, which applies quantum cryptographic techniques to secure electronic voting, to ensure privacy, integrity, and resistance to tampering [30]. Unlike traditional systems vulnerable to cyberattacks, quantum voting promises the use of quantum-secure applications such as QKD to protect vote transmission, making any interception detectable. There are proposals in the literature to potentially use quantum teleportation for this as well [31]. Additionally, blind quantum computation [26] is expected to allow vote processing without revealing individual choices, ensuring both security and anonymity. With further development, quantum networks could enable decentralized and tamper-resistant election verification, reducing risks of fraud and manipulation.

Most of the applications we introduced above require distribution of entangled pairs, which is not an easy technical feat on large distances or scales. In the next subsection, we focus on the concept of repeaters for quantum internet that are essential for achieving this goal.

## 2.5. Repeaters

A complication in the design of quantum internet is the loss of sent photons in glass fiber links. For links with lengths below 50 km this complication is manageable, but for longer distances it surfaces more substantially. A simple solution to this loss is to add to the links repeaters that capture every 50 km the sent photons, for determining their states and for emitting "fresh" photons with the same state along the next 50 km. These repeaters can consist of digital detectors, in which case the repeaters are called digital or "trusted" repeaters.

The disadvantage of these digital repeaters is that they do not maintain the intrinsic security that the links of quantum internet would have. It is therefore possible that the qubit communication passing these digital repeaters is copied in a way that is not detectable by the senders and receivers. The operators of such a repeater could therefore deceive the senders and receivers by manipulating the quantum signals in a way that mimics a secure transmission while secretly extracting information. Since the repeater itself is digital, it can intercept and resend qubits without introducing detectable anomalies in standard eavesdropping tests, effectively allowing undetected surveillance [27]. As already mentioned, digital repeaters are in literature regularly called "trusted repeaters," where the reference to trust should thus be taken as performative rather than descriptive: for maintaining that qubit communication with digital repeaters is secure, one should hold that the operation of these repeaters can be trusted.

A solution to photon loss over long-distances that would maintain the intrinsic security of quantum internet has been discussed in the literature since the late 1990s [32]. This solution is called quantum repeaters and is currently still being developed ([19], p. 94).

These repeaters boost the qubit communication by the quantum-mechanical phenomenon of entanglement swapping [33] in a way that communicants can detect interception through their communication; hence links connected by quantum repeaters introduce the same intrinsic security as single links in quantum internet do.

There are many uncertainties about the intended configuration of quantum internet, for example, whether it should consist mainly of terrestrial elements or be combined with space components. The demonstration of the satellite quantum link between China and Austria in 2017 [12] provided a glimpse into the potential utility of satellite-based quantum communication between end-users on Earth. There are also proposals on how to enable a global quantum network through introduction of quantum repeaters [34]. In either configuration, repeaters are expected to play an essential role in it.

## 3. Trust in technology

### 3.1. Philosophy on trust

Philosophically, trust involves both epistemic and affective conditions: I trust someone because I can predict their motives, beliefs, and behavior, and because I believe they will not harm me. This trust inherently carries a moral dimension, requiring evidence of the other party's moral motivations. The literature defines trust as the willingness to accept vulnerability based on positive expectations of another's behavior [35] and confidence in another's words and actions [36]. Trust is mutual, influencing both parties' judgments and reactions.

Relevant for our case of quantum internet is the extension of the concept of trust of one person in another, to trust of a person, or user, in a technology. Studies in the field of information systems have identified trust as a key determinant of technology adoption and a critical element in understanding user attitudes toward technology (e.g., [37]). Literature distinguishes some specific elements that characterize trust in technology:
    a. *Functionality* denotes the anticipation that a technology possesses the necessary capacity or capability to perform a specified task.
    b. *Helpfulness*, devoid of moral agency and volition (i.e., will), pertains to the intrinsic characteristics of the technology, such as its adequacy and responsiveness, also in emotional terms.
    c. *Reliability* implies the expectation that a technology operates with consistency and predictability.
    d. *Design* encompasses the trust in the individuals who have developed, managed, or overseen the technology, highlighting that trust in technology inherently involves trust in people [38-41].

Trust in technology also involves a moral dimension, as Nickel [42] (see also [43]) has shown in discussing trust in AI in the medical field. Nickel [42] argues for "a stronger notion of trust in AI applications involving giving it *discretionary authority*, a kind of normative authority."

Let us try to apply these generic criteria to the case of new emerging technology such as quantum internet, that is, to a scenario in which the level of uncertainty in whether the technology works as expected is very high: How can we build trust in an emerging technology, especially when its development and impacts remain uncertain? For taking up these questions we adopt the perspective that technologies are not just hardware but embedded in sociotechnical systems which include, in addition to the hardware and the user, also groups of other people as well as organizations engaging with the technologies. Examples of those groups are other users, operators, designers, producers, and maintenance personnel. Examples of organizations are companies, governmental agencies, regulator institutions, and international bodies such as standardization organizations. From a socio-technical systems perspective, an organization – or any part of it – is a network of interconnected sub-systems. These sub-systems include individuals with distinct capabilities collaborating toward shared goals, following structured processes, leveraging technology, operating within a physical environment, and shaped by collective cultural assumptions and norms. At its essence, the socio-technical approach underscores that the design and performance of any organizational system can only be truly understood and improved when the "social" and "technical" dimensions are seamlessly integrated and treated as interdependent elements of a complex whole (see [44-50].

While a comprehensive socio-technical perspective enriches descriptions of technologies from the hardware and user to numerous groups of other people, various organizations, and the social dynamics between them, this paper deliberately narrows its focus to operators and regulatory institutions. Our primary objective is to examine governance, which we consider a foundational aspect of trust in the development and adoption of quantum internet, and operators and regulatory institutions can help in building that trust. The choice to focus on operators and regulators is not accidental. We argue that building trust in the case of quantum internet requires a clear distinction between the roles of operators and regulators.

A second starting point is that we adopt Lewicki and Bunker [51,52] their analysis of trust building in our analysis by its applicability to sociotechnical systems (e.g., [53]) and for bringing in to focus the development of trust with users that have to take place for emerging technologies.

## 3.2. A Lewicky-Bunker model of trust in technology

Lewicki and Bunker's model outlines three stages in the process of trust building within organizations, offering a framework that, while originally designed for interactions between companies and governments, is highly relevant for understanding trust in emerging technologies. By understanding these stages, we can better assess where a particular technology stands in terms of user trust and what steps might be necessary to move towards deeper, more resilient forms of trust.

- The first stage, Calculus-Based Trust (CBT), represents the most fundamental level of trust. At this stage, trust is based on the consistent behavior of an agent – whether that agent is a device, an individual, a company, or a government. For instance, Agent X begins to trust Agent Y by continually assessing Y's actions and noticing a reliable pattern of behavior. In the context of emerging technologies, this might involve users observing that a new technology consistently performs its intended functions without fail. This basic level of trust is crucial for the initial adoption of technology, as it reassures users that the technology works as expected.

- The second stage in the model concerns Knowledge-Based Trust (KBT), which builds on the information gathered about an agent. Here, Agent X's trust in Agent Y deepens as X accumulates enough knowledge to accurately predict Y's actions. This stage of trust is more sophisticated because it involves not just observing consistent behavior but understanding the underlying reasons and patterns behind that behavior. In the realm of emerging technologies, this level of trust might involve users gaining a deeper understanding of how a technology functions, how it provides specific responses, and how it works in various scenarios. This epistemic level of trust is critical as it moves beyond mere confidence in performance to a more informed trust based on insight and understanding.

- Finally, Identification-Based Trust (IBT) represents the highest level of trust in the Lewicki and Bunker model. At this stage, trust is rooted in a deep identification with the values and intentions of the agent. For Agent X to fully trust Agent Y at this level, there must be a strong alignment in their values and goals, leading to mutual appreciation and a reliable bond. In the context of emerging technologies, achieving this kind of trust would mean that users not only understand and predict the technology's behavior but also feel a strong alignment with the principles and purposes that guide its design and function. This level of trust is particularly significant for technologies that play a crucial role in our lives, such as civil aviation, where trust needs to be based on more than just performance – it must also resonate with our ethical and social values.

Let us now consider Agent X, who is deciding whether to trust a particular technology. For more complex technologies, this decision is one about several entities. In addition to Agent X trusting the device itself, they should also trust two types of agents that interact with the device and with X. These agents are operator people who manipulate the device through design, management, and maintenance, and regulators such as governmental agencies, and international organizations that regulate the use and operation of the technology. The key question is: *What conditions must be met for Agent X to trust the device, operators, and regulators?*

To answer this question with the Lewicki and Bunker model, we must distinguish the three levels of trust Agent X can have for the device, for the operators, and for the regulators (see Table 1).

*Table 1 Lewicki and Bunker conditions of trust building in technology, its operators and regulating organizations*

|  | Device | Operator | Regulator |
|---|---|---|---|
| CBT | Consistency of device behavior (no errors, glitches, malfunctions) | Consistency of the operator's behavior (their actions align with their past behavior patterns) | Consistency of the regulator's behavior (their actions align with their past behavior patterns) |
| KBT | Good level of predictability of device behavior (understanding the technical mechanisms of the device) | Good level of predictability of operator's behavior (intentions and actions are clear, transparent; behavior is quite predictable) | Good level of predictability of regulator's behavior (transparent communication, established track record, clear policies and procedures, responsiveness to feedback) |
| IBT | Device behavior is judged to be consistent with human values (values that are significant in that situation) | Operator values are known, shared, and transparent | Regulator values are known, shared, and transparent |

We assume that the highest level of trust – identification-based trust – in the device, the operators, and the regulators may be possible but is rare. For instance, identification-based trust in a device seems to presuppose a level of agency in the device. Some agents may accept that agency for technologies in general, or specific agentive technologies such as AI; other agents may reject such agency for devices. If such an agency is denied, then the best trust to achieve is knowledge-based trust for devices and identification-based trust for their operators and regulating organizations. More qualifications are possible. As soon as the technology itself becomes more difficult to understand, trust in the device may drop to calculus-based. For other technologies operators or regulators may not be that visible, letting users focus on only one of these agents. For instance, pilots – operators of aircraft – are for many users central to trust in aviation. And trust in software may focus on the firms that have developed it.

Let us, for illustrating this approach, further develop the case of civil aviation. In the early days of commercial aviation, trust in air travel may primarily be calculus-based. Passengers developed trust in airlines by observing the consistent and safe operation of flights over time. This basic trust was built on the reliability of the aircraft, the competence of the pilots, and the overall safety record of the industry. Passengers could see that flights generally arrived on time and that accidents, while they did occur, were rare. This consistency helped establish a foundational level of trust that encouraged more people to consider air travel as a viable option. As commercial aviation developed and became more widespread, passengers' trust in air travel deepened into the knowledge-based level. This shift occurred as people gained more understanding of how airplanes worked, the rigorous training that pilots and crew undergo, and the safety measures in place, such as maintenance protocols and air traffic control systems. Media coverage, public reports on safety, and transparency from airlines contributed to a more informed trust. Today, for many frequent flyers, trust in air travel has reached the highest level: identification-based trust. This level of trust is evident in the way passengers not only rely on air travel for their personal and professional needs but also align with the values and missions of certain airlines. For example, a traveler might choose to fly with a specific airline not just because it is safe and reliable but because they identify with the airline's commitment to environmental sustainability, customer service, or innovation. This identification-based trust is also reflected in the loyalty programs that airlines offer, where customers feel a strong sense of connection and commitment to the brand, beyond mere functional reliability.

Having spelled out how we approach trust in emerging technology, we return to our case of quantum internet.

# 4. Trust in Quantum Internet

## 4.1. Trusting repeaters

A first observation is that trust building in quantum internet equipped with digital trusted repeaters aligns closely with that in the existing digital internet. In such networks, the trustworthiness that communication sent through the infrastructure is secure, largely hinges on the credibility and reliability of the operators and the organizations that manage them. The same applies to quantum internet with trusted repeaters: as a user you are monitoring if the devices display their desired behavior, which would be calculus-based trust in the Lewicki and Bunker model. Yet the primary concern for users remains whether the operators and their organizations can be trusted not to misuse their control over the data flow, and how these operators are guided and overseen. Users can develop higher levels of trust in these operators and regulators, which compensates possible epistemic constraints of users to fully understand and check hardware. We contend that trust in quantum internet equipped with trusted repeaters must be established in a similar manner: users can arrive at trust through the trusting of the operators and regulators.

The above observation seems not to hold when quantum internet can be equipped with quantum repeaters. Communication seems then to be secure independently of what operators and regulators do, and intrinsically secure communication may come available through just the hardware and software of quantum internet. The quantum repeaters are doing by the laws of quantum mechanics what they are designed for, which would enable in the end identification-based trust on these devices, without a need to also trust the operators and regulators. Yet, a crucial part of this shift relies on the end user's capability to verify that they in fact share an exclusive correlation with their intended communication partner in a device independent manner. A recent technical report by *The Federal Office for Information Security* (BSI) in Germany [18] gives evidence that users may lack this capability: the communicating partners can be fooled.

The BSI report extensively analyzes the technical vulnerabilities of QKD systems aimed at enabling senders and receivers to create shared secret keys.[2] The main focus is on protocols that rely on direct qubit transmission (such as BB84) but the report also covers protocols that use entanglement distribution (E91). For example, Table 4.43 in the BSI report ([18], p. 116]) mentions imperfections in the quantum channels that might give eavesdroppers the technical possibilities of getting partial access to the correlations between end-nodes created by entanglement. Similarly, Table 4.62 ([18], p. 144) describes an attack on the sensors used in protocols with entanglement distribution for tricking the sensors into giving out "clearance" data by which senders and receivers believe that the system is free from eavesdropping on, while in fact this data is caused by the manipulation of the sensors.

In the next subsection we argue that these results about ways to fool senders and receivers into believing that they exchanged encryption keys security brings trust building in the quantum internet again back to its operators and regulators.

## 4.2.  Trusting communication

The ideal for quantum repeaters is that trust that relayed communication is not intercepted, copied, or manipulated can depend on the technical structure of quantum internet rather than on its operators and regulators. The BSI report [18] undermines this possibility, as it shows that users cannot easily determine on the basis of their interactions with the devices that they are dealing with a secure infrastructure. Calculus-based trust on the level of devices is therefore ruled out, since the behavior of the devices do not reveal whether quantum internet is providing secure communication or is under attack. Knowledge-based trust via constantly studying the potential vulnerabilities of the systems to predict how they will behave under certain attacks provides a better alternative. Furthermore, it also again brings the question of trust back to the operator and regulating organizational levels. Due to the existence of potential attacks, even if the systems are theoretically 100% secure under theoretical conditions, the basis for users to trust the infrastructure enough to adopt it for high-value information exchanges, will become trust in the ability of the operators of quantum internet and the capabilities and intentions of the organizations to monitor attacks and to warn the users about them.

---

[2] The BSI report [18] looks at *implementation attacks* on quantum internet used for QKD, which are attacks on the technical elements of quantum internet rather than attacks that attempt to just incept the sent data.

All these considered, for the case of digital and quantum repeaters we can argue that a quantum network that relies on digital repeaters as trusted relays provides no real change to the trust dynamics compared to the digital internet. It just shifts and concentrates vulnerabilities to the operation of the trusted repeater nodes. In contrast, a quantum network empowered by quantum repeaters initially shifts the trust dynamics more towards the device level. Yet the operator and organization levels are still in play through the question of whether the entanglement distribution section of the infrastructure is protected against attacks or not. Introduction of entanglement as a resource to the system shifts the trust dynamics to the implementational fidelity on the quantum network side.

We are not the first ones to highlight this point, for example, in his book *Schrödinger's Web: Race to Build the Quantum Internet* Jonathan Dowling argues that

> Someday there will be a Department of Entanglement whose job it is to keep the entanglement coming just like the Department of Energy has the job to keep the lights on. We can imagine millions or billions of these little entanglement-spewing machines all over the network – constantly providing the users with shared entanglement. ([54], p. 132)

Distribution of entanglement forms the basis of quantum internet, and governance of this distribution forms the basis of trust into this technology. Hence for developing trust in quantum internet, the focus should be on creating high levels of trust in the operators and organizations running it. And, taking again distance from Dowling, in these geopolitically more complex times, a Department of Entanglement from one national state (or one Tech company) will not do. Users may arrive at calculus-based or possibly even knowledge-based trust in single national organizations or commercial companies; for identification-based trust, more independent supranational organizations are needed. We further develop these points by moving to the governance of quantum internet.

## 4.3. Trusting governance

Full protection of communication is compelling by the protection of privacy it brings, yet we typically also want a fair degree of governance of this protection of communication [55]. Protection of one's communication provides personal security yet can lead in societies to nonsecure phenomena, such as the untransparent and criminal forms of communication we currently see in the dark net within the digital internet. This dilemma brings to the forefront critical issues of governance of quantum internet: determining who can infringe upon one's privacy and under what circumstances.

Approaches to dealing with this dilemma can draw from discussions on the regulation of the digital internet. And this is still an open issue. Looking at the Snowden revelations bringing forth the methods of massive surveillance by states into question [56], at the US Senate discussing banning TikTok due to potential foreign interference [57], at the arrest of Telegram's CEO for failing to prevent use that platform for criminal activity [58], and at the EU's so called "Chat Control" proposal to end privacy for all digital correspondence taking place in Europe [59]; what can be concluded is that there is still no political or societal consensus about what the right balance of privacy and security is in the digital realm, and about who should play which role in maintaining it. To navigate these waters for our case of quantum internet, strategic implementation of rules related to the governance of quantum internet at various levels – technological, operational, and legal – is crucial. These rules must not only promise trust but ensure it is tangibly experienced by users, thereby fostering an environment where quantum internet can be socially accepted and utilized without apprehensions.

Establishing trust in quantum internet necessitates a thoughtful approach to regulation. Initially, trust might be calculus-based, which hinges on the consistent and predictable behavior of the technology. Here, rules mandating transparency and consistency in operations can serve as the foundation for building user trust. However, as familiarity with quantum internet grows and its adoption increases, moving towards knowledge-based and identification-based trust becomes essential. These higher levels of trust are directed to the operators of quantum internet and to their regulatory bodies. These layers must work in concert to ensure that the technology not only adheres to societal norms and expectations but also resonates with societal values, thereby deepening trust. To facilitate this, regulatory frameworks can be designed to be flexible yet robust enough to adapt to the evolving landscape of quantum technology. This includes establishing clear guidelines for data protection, user privacy, and technology deployment that are comprehensible to all stakeholders involved – from tech developers to end-users.

In practical terms, this requires strong oversight. Just because there are clear guidelines does not mean they will be followed, and for this regulatory approach to work there needs to be oversight mechanisms that can track whether the guidelines are being followed by all the stakeholders or not. One argument from the Snowden files was that the oversight mechanisms in place to prevent the establishment of mass surveillance tools, such as the United States Foreign Intelligence Surveillance Court (FISA) in the NSA case, were rendered useless after the shock of 9/11 terror attacks against the US. Out of 33942 warrants to conduct surveillance activities, the court denied only 12 – a rejection rate of just 0.03%. Similarly, another oversight mechanism in the U.S., being Senate hearings, was essentially rendered ineffective when the director of national intelligence falsely testified and refused the existence of such programs for mass surveillance. It was a clear demonstration of the inadequacy of public hearings as it became apparent that civil servants do not refrain from lying under oath and misinforming the public in the name of other concerns, such as national security. Today, we know that those tools and programs exist [60], and they even form the basis of concerns over *harvest now-decrypt later* attacks using quantum computers for digital applications such as e-voting ([61], pp. 21-22).

To avoid that quantum internet will be subject to oversight by national actors due to their national interests, an obvious first choice would be to propose the implementation of international oversight mechanisms. However, this is not a straightforward or easy to achieve aim. The inherent complexity and vast potential of quantum internet make it particularly susceptible to the risks associated with overregulation. Excessive regulatory measures, while intending to enable trust and security, might constrain the technology's inherent capabilities and flexibility. The challenge lies in crafting regulations that establish a baseline of trust without stifling innovation and convince the national actors to moderate demands on maximizing their national interests.

To mitigate the risk of overregulation, integrating design choices that inherently enhance transparency and accountability can prove beneficial. Such measures not only align with the advanced stages of trust development (knowledge and identification-based) but also empower users by providing clear insights into the technology's operation and governance. This encourages a more informed and voluntary trust, moving beyond mere compliance to genuine understanding and appreciation of the technology behind quantum internet. Moreover, concepts such as "privacy by design" can be pivotal here, embedding privacy considerations into the very fabric of quantum internet and by proactively addressing potential privacy issues through technology design, rather than retroactively through regulatory measures. This can prevent the stifling effects of overregulation while ensuring a secure and trustworthy quantum internet.

Therefore, national dominance in overseeing the quantum internet risks reducing the number of independent trust sources available to users. According to the Lewicki and Bunker model, sources of trust can come from the technology, its operators, and its regulators. If Big Tech controls both development and operation, or if states nationalize both operation and regulation, these sources collapse into one. To preserve user trust, it is crucial to maintain separation and independence among these domains.

# 5. On building trust

This paper challenges the assumption that quantum internet can deliver inherently secure communication solely through its technical design. Instead quantum internet can fulfill its promises if the policies and frameworks for running it can earn public trust. Within the Lewicky and Bunker model this trust consists, as said, of three elements: trust in the hardware and software of quantum internet, trust in the operators that make quantum internet run, and trust in the regulators that oversee quantum internet and its operators. Trust in the hardware and software is for a typical user hard to create since it involves extensive knowledge of the hardware and operation of quantum internet as well as the ability to apply tests and protocols for checking if it runs as expected; for users trust in the operators and the regulators themselves is therefore a good or sometimes sole alternative.

Building a quantum internet that can be trusted thus involves creating also the operating and regulation of this infrastructure. Sketching the contours of this governance of quantum internet lies beyond our expertise, which is in philosophy of technology and technology assessment. Yet we can offer to decision-makers involved in the development of quantum internet and of its management ingredients to governance that may lead to a quantum internet that will be trusted.

## 5.1. *Stakeholder engagement*

Reaching out to the stakeholders involved is a standard approach within technology ethics for learning from users and society at large what they think of, expect from, and (dis)value in a technology, and for involving them in the development of a technology. This stakeholder engagement can be of value to quantum internet too.

Trust is not a static attribute; it is continuously shaped by interactions between diverse stakeholders, including governments, industry leaders, academics, civil society, and end-users. Ensuring meaningful stakeholder engagement is, therefore, a crucial aspect of governance. One key ethical governance approach is stakeholder mapping, which systematically identifies the actors influencing or affected by quantum internet development. A method such as *value-sensitive design* (VSD) can help integrate the diverse values and concerns of these stakeholders into the design and operationalization of quantum internet [62]. This ensures that principles such as equity, accessibility, and security are embedded into the infrastructure from the outset [63]. Furthermore, an inclusive collaboration model is necessary to address disparities between different stakeholder groups, particularly between technologically advanced nations and those with less access to cutting-edge research. Transparent governance processes must bridge these gaps, ensuring that quantum internet does not exacerbate digital divides but instead benefits all sectors of society [64,65].

Since trust is not a one-time achievement but an evolving relationship, long-term trust maintenance strategies are needed. Governance must anticipate and address emerging vulnerabilities, such as new cryptographic risks or unforeseen ethical dilemmas. Historical trust dilemmas in technology adoption [66,67] illustrate that without proactive governance, initial trust can erode due to perceived or real threats. Quantum internet's governance model must incorporate continuous risk assessment and adaptive policy frameworks to sustain trust throughout its lifecycle. The co-evolution of quantum technology and governance is essential – the success of quantum internet depends on its integration into existing societal infrastructures while shaping new norms and institutions [68]. Governance must ensure that operators, policymakers, and institutions are equipped to manage these transitions responsibly [64,65,68].

## 5.2. Recommendations

More specific considerations that we deduce from our analysis are the following.

- **Role diversification**
  In our analysis we approached quantum internet from a sociotechnical perspective and distinguished its hardware and software from its operation and regulation. For users all these three elements are sources for building trust in quantum internet. For supporting users in having these multiple sources for trust, it is recommended to avoid governance models of quantum internet in which single organizations control two or three of these elements. Ideally independent companies and organizations are involved in delivering and managing these elements. And when, say, hardware, software, and daily operator services are delivered by one company, the recommendation is to implement regulation in a second independent organization. Commercial developments and geopolitical interests may result in monopolies by Big Tech companies or national states, which are better avoided when building a quantum internet that is to be trusted.

- **User checks**
  Checking the hardware and software is, as noted in our analysis, typically hard to do for typical users, which they can compensate for by trusting the operators and regulators of quantum internet. An alternative mechanism for users to trust hardware of a technology is third party organizations – let us call them civil interests organizations – that can do the checking for them. One example of this in recent German history is the civil activism of organizations like the Chaos Computer Club [69] that resulted in the ban of e-voting machines in German elections by the Federal Constitutional Court [70]. We recommend that such civil-interest organizations are added to the sociotechnical fabric of quantum internet.

- **Global interoperability**

  For preventing technological fragmentation and inequitable access, international collaboration is essential in establishing standardized quantum communication protocols. A failure to harmonize these standards could lead to a fractured quantum internet where certain regions or actors dominate, limiting its democratizing potential and limiting the ability of multinational companies to use quantum internet to secure their international communication.

- **Identification-based trust in surveillance**

  Trust in regulation of quantum internet will involve trust in surveillance and policing of usage. Full transparency of this aspect in terms of openness about, say, all rules under which governmental organization can intervene, would establish calculus- and knowledge-based trust in the regulation. Such trust seems unrealistic by the inherent covertness of surveillance and policing. Still this recognition should not block efforts of aiming at trust in regulation. It can be argued that public deceptions about surveillance of usage of digital internet – think about the massive collection of personal data by Big Tech and US governmental agencies – have boosted the development of quantum internet for the security of communication it may offer. Similar deceptions should be avoided for quantum internet and may be realized by aiming at identification-based trust of users in the regulation of quantum internet.

# 6. Conclusion

In this paper, we have analyzed the social conditions for the usage of quantum internet for its promise to deliver intrinsically secure communication by its technical structure. We have argued against this promise by showing that the trust that users can build up in secure communication by quantum internet will also depend on the way in which quantum internet is operated and regulated. Central in this argument is the recognition that quantum internet can be attacked by third parties for letting users believe that communication is secure.

Trust is the cornerstone of not only adopting new technology but also in ensuring its successful integration into society. The relationship between social acceptance and technological innovation is intricate and reciprocal. Innovations must be socially accepted to be widely adopted, yet social acceptance itself hinges on a clear demonstration of the innovation's benefits and its alignment with societal values. For quantum internet the relevant values may be privacy and security. As such, social acceptance of quantum internet will largely depend on how these values are balanced: the public's trust in quantum internet and their perceptions of the technology's benefits versus its risks, will become deciding factors in the adoption or rejection of this new communication technology.

# Acknowledgements

Work by Pieter E. Vermaas was supported by the Dutch National Growth Fund (NGF), as part of the Quantum Delta NL programme. Luca Possati acknowledges the support of the Sector Plan for Science and Technology Education (Netherlands). Zeki C. Seskir acknowledges support by the European Union under the IANUS ("INspiring and ANchoring TrUst in Science") grant (Grant Agreement ID: 101058158). The views and opinions expressed are those of the author(s) alone and do not necessarily reflect the official position of the European Union or the European Research Executive Agency.